\documentclass[conference]{IEEEtran}

\usepackage{amsmath,amssymb}
\usepackage{balance}
\usepackage{booktabs}
\usepackage{cite}
\usepackage{graphicx}
\usepackage[hidelinks]{hyperref}
\hypersetup{
  pdfauthor={Chi-Sheng Chen, Yun-Wen Chen},
  pdftitle={Global Response Normalization in a Convolutional Transformer Network for Cross-Session Motor Imagery EEG}
}

\newcommand{\pp}{\,\mathrm{pp}}
\newcommand{\Drop}{\operatorname{Drop}}
\newcommand{\vecop}{\operatorname{vec}}
\newcommand{\range}{\operatorname{span}}

\begin{document}

\title{Global Response Normalization\protect\\
in a Convolutional Transformer Network\protect\\
for Cross-Session Motor Imagery EEG}

\author{%
\IEEEauthorblockN{Chi-Sheng Chen}
\IEEEauthorblockA{Brain Research Center, National Tsing Hua University\\
Hsinchu, Taiwan\\
Quantquer Inc., Taoyuan, Taiwan\\
\href{mailto:m50816m50816@gmail.com}{m50816m50816@gmail.com}}
\and
\IEEEauthorblockN{Yun-Wen Chen}
\IEEEauthorblockA{Department of Management Information Systems\\
National Pingtung University of Science and Technology\\
Pingtung, Taiwan\\
\href{mailto:yunwenchen86@gmail.com}{yunwenchen86@gmail.com}}}
\maketitle

\begin{abstract}
Trial-dependent feature gains could help motor imagery EEG decoders transfer across sessions. We add global response normalization (GRN) to the convolutional transformer network (CTNet) and compare it with the original backbone and a 32-parameter affine control. With nine participants per dataset and five paired seeds, GRN changed strict accuracy relative to CTNet by $+0.25\pp$ on BCI Competition IV-2a (95\% CI $[-0.06,+0.56]\pp$, $p=0.191$) and $-0.02\pp$ on IV-2b ($[-0.21,+0.15]\pp$, $p=0.859$). IV-2a CTNet scores varied from 76.47\% to 81.71\% across protocols. With trained weights fixed, replacing the response by its fitting-set mean changed 28 of 12,960 top-1 decisions on IV-2a and 13 of 14,200 on IV-2b. An energy-weighted response lowered fitting-set feature error and produced more evaluation flips on both datasets. The paired comparisons found no detectable accuracy gain from this GRN bridge.
\end{abstract}


\section{Introduction}

Motor imagery brain--computer interfaces decode imagined movements from electroencephalography (EEG). A participant-specific decoder can lose accuracy on a different recording day \cite{cross_session}. EEGNet learns temporal and spatial filters, and QEEGNet examines encoder changes on IV-2a and IV-2b \cite{eegnet,qeegnet}. EEG Conformer \cite{conformer} and the convolutional transformer network (CTNet) \cite{ctnet} use self-attention to integrate trial features. Comparisons of covariance embeddings in a shared EEG Transformer illustrate how a fixed backbone can isolate representation choices \cite{spd}.

ConvNeXt V2 introduced global response normalization (GRN) to encourage feature competition in masked image pretraining \cite{convnextv2}. The MAEConformer preprint uses temporal GRN in place of batch normalization for neonatal EEG \cite{maeconformer}. We hypothesize that trial-dependent gains can help preserve informative features across motor imagery sessions. Temporal GRN pools responses over the trial and normalizes them by their mean across feature channels. The affine control has the same per-channel scale and bias parameters and uses a fixed response.

We first measure the paired accuracy effect of adding GRN under strict validation on IV-2a and IV-2b, and use the affine comparison to assess the contribution of trial dependence. Additional IV-2a protocols show how the comparison changes with fitting and checkpoint selection. We then freeze the response within each trained GRN model to measure its contribution to inference. We derive a feature-error minimizer and compare both fixed responses using observed decision flips and a logit-margin diagnostic. Electrode perturbations and masked reconstruction probe specific sensor changes and a short pretraining schedule.

\section{Models and Evaluation Protocols}

\begin{figure*}[!t]
  \centering
  \includegraphics[width=\textwidth]{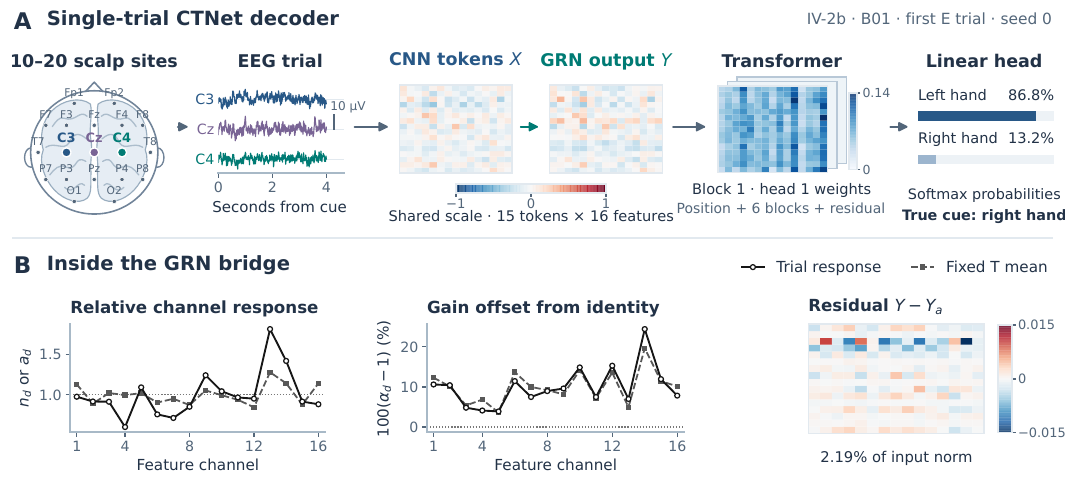}
  \caption{CTNet on B01's first IV-2b evaluation trial (strict GRN, seed 0). (A) Nominal bipolar sites C3, Cz, and C4 on the 10--20 map \cite{bcic2b}, offset EEG traces, tokens (feature rows, time columns), attention (query rows, key columns), class probabilities and true cue. (B) Native and fixed T-mean responses, gains $\alpha_d=1+\gamma_d n_d$ or $1+\gamma_d a_d$, and the feature residual.}
  \label{fig:architecture}
\end{figure*}

\subsection{Datasets and CTNet Backbone}

BCI Competition IV-2a contains nine participants with two sessions of 288 trials each, 22 EEG channels and four motor imagery classes (left hand, right hand, feet and tongue) \cite{bcic2a}. IV-2b contains nine participants, three bipolar EEG channels (C3, Cz and C4) and two hand imagery classes \cite{bcic2b}. In IV-2b, training data (T) pool two screening sessions and one feedback session (400--440 trials), and evaluation data (E) pool two feedback sessions (280--320 trials). Offline decoding uses the first four seconds from each cue at 250 Hz. We exclude EOG and preserve acquisition filtering (0.5--100 Hz bandpass and 50 Hz notch). The primary analyses retain artifact-marked trials. We train one decoder per participant and fit a scalar mean and standard deviation on its fitting trials.

The shared convolutional stem uses eight temporal filters of length 64, depthwise spatial filtering over electrodes with multiplier two, and a second temporal convolution of length 16. Two average pooling layers with stride eight yield $L=15$ tokens of width $D=16$. For token matrix $X\in\mathbb R^{L\times D}$, CTNet with bridge $\mathcal B$ computes
\begin{equation}
  \begin{aligned}
    z &= \Drop_{0.1}\!\left(\sqrt D\,\mathcal B(X)+P_{1:L}\right), \\
    o &= W\Drop_{0.5}\!\left(\vecop[z+\mathcal T(z)]\right)+b.
  \end{aligned}
  \label{eq:ctnet}
\end{equation}
Here $P_{1:L}$ gives the learned positional embeddings and $\mathcal T$ is a stack of six two-head, post-normalized Transformer blocks. The readout $W$ maps 240 features to four or two logits, depending on the dataset. The identity bridge reproduces upstream checkpoint logits on both datasets. Fig.~\ref{fig:architecture} shows the bridge before positional encoding.

\subsection{GRN and Parameter-Matched Control}

For a single trial, the bridge computes
\begin{align}
  g_d(X)&=\|X_{:d}\|_2,\qquad \bar g(X)=\frac{1}{D}\sum_{j=1}^{D}g_j(X),\\
  n_d(X)&=\frac{g_d(X)}{\bar g(X)+10^{-6}},\label{eq:response}\\
  \mathcal B(X)_{\ell d}&=X_{\ell d}+\gamma_d X_{\ell d}n_d(X)+\beta_d.\label{eq:grn}
\end{align}
The gain is shared over the $L$ tokens within each trial. The affine control fixes $n_d(X)=1$ and has the same gain and bias parameters. Both bridges add $2D=32$ parameters and initialize $\gamma$ and $\beta$ to zero. The baseline, GRN and affine models share initial backbone weights and initial predictions for each participant and seed.

\subsection{Training and Statistical Analysis}

The strict protocol fixes a class-stratified T split for each participant and seed (0--4), shared by all variants. IV-2a uses 204 fitting and 84 validation trials. IV-2b holds out 30\% of each class, rounded down, leaving 280--308 fitting trials. We fit the standardizer and draw augmentation from fitting trials. Each synthetic trial combines eight time-aligned segments from the same class. Training uses Adam ($10^{-3}$, $\beta_1=0.5$, $\beta_2=0.999$) and cross-entropy for 1,000 epochs, with 216 reconstructed trials per update. IV-2a has four updates per epoch and IV-2b has six or seven. Variants share backbone initialization, batch order, augmentation draws and dropout streams. We select the first T-validation loss minimum, then evaluate on E.

As a secondary IV-2a comparison, we reproduce the upstream protocol with rotating validation. Each epoch reshuffles T into four groups of 72, holding out 21 trials per group for that epoch. Trials can contribute directly to fitting in other epochs, and the standardizer and reconstruction pool use all of T. Strict and rotating validation both use 51 recorded trials per update. E is held out for final evaluation.

In the IV-2a full-T follow-up, each model is retrained on all 288 T trials for the duration selected by its strict run, with 72 recorded and 216 reconstructed trials per update. Standardization and augmentation use all of T. The GRN arm and decision rule were specified before its 45 refits, with the matched CTNet refit results already known.

We average paired accuracy differences over seeds within participant, then average the nine participant means per dataset. Two-sided sign-flip tests enumerate all $2^9$ assignments under a sign-exchangeable null. Separately, we obtain pointwise 95\% percentile bootstrap intervals (CIs) from 20,000 participant resamples. The strict and full-T GRN--CTNet tests require $p<0.05$ and a positive lower CI bound. The IV-2b analysis plan was recorded internally before training. Analyses introduced after observing E are descriptive.

\section{Fixed-Response Analysis}

\begin{figure*}[!t]
  \centering
  \includegraphics[width=\textwidth]{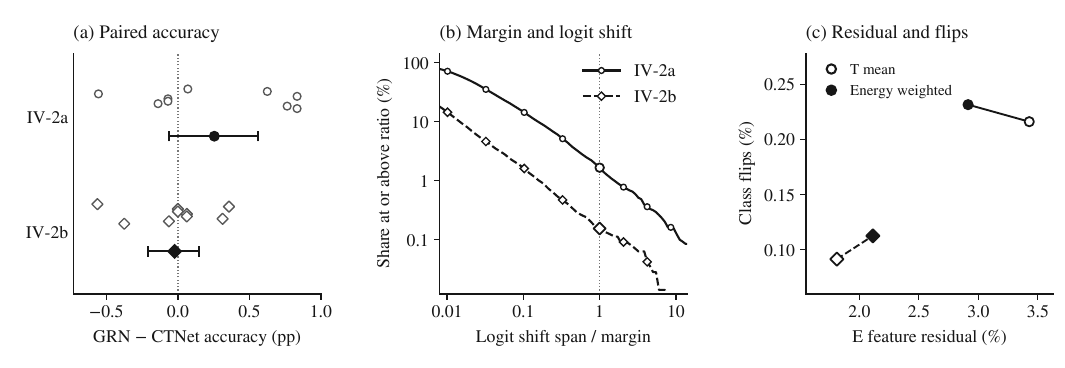}
  \caption{Accuracy and fixed-response replay. (a) Participant means over five seeds (open) and the overall mean with a 95\% participant-bootstrap interval (filled). (b) Empirical tail proportions of $\range(\delta)/m$ for the T mean. Lines connect evaluated thresholds, with a dotted line at one. (c) T mean (open) and energy-weighted mean (filled). Circles denote IV-2a and diamonds denote IV-2b.}
  \label{fig:effects}
\end{figure*}

\subsection{Affine Approximation of the Bridge}

Write $Y(X)=\mathcal B(X)$ for a trained GRN bridge. For a fixed channel vector $a$, let $Y_a$ be the bridge with $n(X)$ replaced by $a$. The channel-wise residual between these two maps is exactly
\begin{equation}
  R_a(X)_{\ell d}=Y(X)_{\ell d}-Y_a(X)_{\ell d}
  =\gamma_d X_{\ell d}[n_d(X)-a_d].
  \label{eq:residual}
\end{equation}
The fixed map has scale $1+\gamma_d a_d$ and bias $\beta_d$. Summing over the fitting set $T_{\mathrm{fit}}$ gives
\begin{equation}
  \sum_{i\in T_{\mathrm{fit}}}\|R_a(X_i)\|_F^2
  = \sum_{i\in T_{\mathrm{fit}}}\sum_d
    \gamma_d^2 g_d(X_i)^2[n_d(X_i)-a_d]^2.
  \label{eq:residual_loss}
\end{equation}
For a channel with nonzero gain and positive total energy, setting the derivative with respect to $a_d$ to zero gives the unique minimizer
\begin{equation}
  a_d^\star = \frac{\sum_{i\in T_{\mathrm{fit}}} g_d(X_i)^2 n_d(X_i)}
  {\sum_{i\in T_{\mathrm{fit}}} g_d(X_i)^2}.
  \label{eq:weighted}
\end{equation}
Any constant minimizes the fitting error if either the gain or the total channel energy is zero. Equation~\eqref{eq:weighted} minimizes feature error on the fitting data with $\gamma$ and $\beta$ fixed. An arbitrary constant $a_d$ incurs excess squared error $\gamma_d^2 (a_d-a_d^\star)^2\sum_i g_d(X_i)^2$ in that channel, as the weighted deviations about $a_d^\star$ sum to zero. Across sessions, a different distribution of channel energies and responses can favor a different constant.

\subsection{Logit-Margin Diagnostic}

Let $o$ be the native logit vector, $\tilde{o}$ the fixed-response logits and $c$ the native winning class. Define $m=o_c-\max_{j\ne c}o_j$ and $\delta=\tilde{o}-o$. For each competing class $j$,
\begin{equation}
  \tilde{o}_c-\tilde{o}_j
  \;\ge\;m-\range(\delta),\quad
  \range(\delta)=\max_k\delta_k-\min_k\delta_k.
  \label{eq:margin}
\end{equation}
Thus $m>\range(\delta)$ is sufficient for an unchanged prediction in the replay. A shift common to all logits cancels. The fraction outside this condition upper-bounds the observed flip fraction. With equal decision weights it also bounds the absolute change in pooled accuracy. Computing $\delta$ requires both forward passes, so this check applies to the observed replays. We use actual flips and participant-level accuracy changes to quantify the intervention.

For each strict GRN checkpoint, we compute $a$ in evaluation mode on only its fitting trials. We compare the ordinary mean response and $a^\star$ in~\eqref{eq:weighted}. The native and fixed-response replays share the backbone, $\gamma$ and $\beta$ and use CPU batches of 72. For each data split we compute $\|R_a\|_F/\|X\|_F$ pooling trials, tokens and channels within a checkpoint, then average the checkpoint ratios. Zero-margin decisions fall outside the sufficient condition.

\section{Results}

\subsection{Paired Accuracy Across Protocols}

\begin{table}[!b]
  \caption{Cross-session accuracy and paired GRN effects by protocol}
  \label{tab:main}
  \centering
  \footnotesize
  \setlength{\tabcolsep}{2.8pt}
  \begin{tabular}{@{}lrrrr@{}}
    \toprule
    Protocol & CTNet & +GRN & $\Delta$ [95\% CI] & $p$ \\
    \midrule
    IV-2a strict & 76.47 & 76.72 & $+0.25\ [-0.06,+0.56]$ & 0.191 \\
    IV-2a rotating & 81.71 & 81.40 & $-0.31\ [-0.81,+0.15]$ & 0.293 \\
    IV-2a refit  & 79.30 & 79.36 & $+0.06\ [-0.31,+0.33]$ & 0.816 \\
    IV-2b strict & 85.72 & 85.70 & $-0.02\ [-0.21,+0.15]$ & 0.859 \\
    \bottomrule
  \end{tabular}
  \par\smallskip
  \parbox{\columnwidth}{\footnotesize Accuracies are single-model percentages averaged across seeds. Effects and 95\% CIs use unrounded scores and are in percentage points.}
\end{table}

Neither strict comparison met the improvement rule (Table~\ref{tab:main}, Fig.~\ref{fig:effects}(a)). GRN minus affine was $+0.12\pp$ on IV-2a (95\% CI $[-0.17,+0.42]\pp$, raw $p=0.488$) and $+0.14\pp$ on IV-2b ($[-0.15,+0.42]\pp$, raw $p=0.406$). The IV-2a affine contrast had Holm-adjusted $p=1.000$ over six exploratory comparisons \cite{holm}.

On retrospective scoring of raw E softmax probabilities (270 runs), GRN minus affine log loss was $+0.0050$ nats on IV-2a and $+0.0005$ on IV-2b. For log loss and multiclass Brier score \cite{scoring}, all eight contrasts against CTNet or affine had Holm-adjusted $p=1.00$. The archive reports both scores and intervals \cite{materials}.

Retrospective scoring on E trials without artifact flags (event 1023) retained 2,368 of 2,592 trials on IV-2a and 2,241 of 2,840 on IV-2b. GRN minus CTNet was $+0.34\pp$ on IV-2a (95\% CI $[+0.04,+0.62]\pp$, $p=0.0625$) and $+0.02\pp$ on IV-2b ($[-0.21,+0.22]\pp$, $p=0.891$). GRN minus affine was $+0.17\pp$ and $+0.18\pp$, respectively.

With rotating validation, CTNet gained $5.25\pp$ over strict training. Overlap within T coincided with 100\% median validation accuracy and median selection at epoch 949 for CTNet and 941 for GRN. Full-T refit raised CTNet accuracy by $2.83\pp$ with the epoch count set by strict validation. The cross-protocol differences combine changes to the fitting data, augmentation pool and selection rule. Each GRN effect in Table~\ref{tab:main} pairs variants within its protocol.

All 90 refit checkpoints reproduced their archived E logits. Retrospective averaging of five strict checkpoints' E probabilities raised CTNet accuracy to 81.37\% on IV-2a ($+4.90\pp$) and 86.95\% on IV-2b ($+1.23\pp$, $p=0.078$). The IV-2b GRN ensemble scored $0.13\pp$ below both CTNet ($p=0.438$) and affine ($p=0.469$) \cite{materials}.

\subsection{Contribution of the Trial-Dependent Response}

\begin{table}[!t]
  \caption{Feature residuals and decision changes after freezing GRN}
  \label{tab:margin}
  \centering
  \footnotesize
  \setlength{\tabcolsep}{3.2pt}
  \begin{tabular}{@{}llrrrrr@{}}
    \toprule
    & & \multicolumn{2}{c}{Residual (\%)} & & & \\
    \cmidrule(lr){3-4}
    Data & Response & T-fit & E & \shortstack{Margin condition\\met (\%)} & Flips & $\Delta$ (pp) \\
    \midrule
    IV-2a & Mean     & 2.75 & 3.43 & 98.33 & 28 & $+0.023$ \\
          & Weighted & 2.29 & 2.91 & 98.57 & 30 & $+0.085$ \\
    \addlinespace[2pt]
    IV-2b & Mean     & 3.07 & 1.81 & 99.85 & 13 & $+0.021$ \\
          & Weighted & 2.06 & 2.11 & 99.80 & 16 & $+0.042$ \\
    \bottomrule
  \end{tabular}
  \par\smallskip
  \parbox{\columnwidth}{\footnotesize T-fit denotes the strict fitting subset. The margin condition is $m>\range(\delta)$. Flips count observed decision changes, pooled over 12,960 IV-2a and 14,200 IV-2b trial--seed decisions. $\Delta$ is native minus fixed accuracy, averaged over seeds within participant.}
\end{table}

The T-mean replacement changed 28 of 12,960 decisions on IV-2a and 13 of 14,200 on IV-2b (Table~\ref{tab:margin}). Native minus fixed accuracy was $+0.023\pp$ (95\% CI $[-0.031,+0.085]\pp$) and $+0.021\pp$ ($[-0.035,+0.076]\pp$), respectively. The margin condition held for 98.33\% and 99.85\% (Fig.~\ref{fig:effects}(b)). Native CPU replay matched all 27,160 archived GPU decisions.

Energy weighting lowered the fitting-set residual on both datasets. On E the IV-2a residual fell from 3.43\% to 2.91\%, with flips rising from 28 to 30 (Fig.~\ref{fig:effects}(c)). Both increased on IV-2b. Native minus weighted accuracy was $+0.085\pp$ on IV-2a (95\% CI $[-0.023,+0.216]\pp$) and $+0.042\pp$ on IV-2b ($[-0.007,+0.097]\pp$).

\balance
\subsection{Sensor Perturbations on IV-2a}

To probe the session-variation hypothesis, we replayed the strict CTNet, GRN and affine checkpoints with three shared perturbation draws per participant. We applied log-uniform electrode gains before standardization and simulated missing electrodes by setting them to zero after standardization. Each draw was held constant across a participant's E trials and paired across seeds and variants. Retention is the change from each model's own clean accuracy.

\begin{table}[!b]
  \caption{IV-2a retention relative to clean replay (pp)}
  \label{tab:stress}
  \centering
  \footnotesize
  \setlength{\tabcolsep}{2.6pt}
  \begin{tabular}{@{}lrrrr@{}}
    \toprule
    Perturbation & GRN & Affine & $\Delta$ [95\% CI] & Raw $p$ \\
    \midrule
    Gain $0.5$--$2\times$  & $-34.93$ & $-34.78$ & $-0.15\ [-0.62,+0.32]$ & 0.574 \\
    Gain $0.25$--$4\times$ & $-45.52$ & $-45.21$ & $-0.31\ [-0.56,-0.09]$ & 0.031 \\
    Two missing           & $-27.43$ & $-26.78$ & $-0.64\ [-1.63,+0.11]$ & 0.242 \\
    Five missing          & $-40.41$ & $-39.88$ & $-0.52\ [-1.18,+0.03]$ & 0.176 \\
    \bottomrule
  \end{tabular}
  \par\smallskip
  \parbox{\columnwidth}{\footnotesize $\Delta$ is GRN minus affine retention, computed before rounding.}
\end{table}

All four GRN-minus-affine estimates were negative (Table~\ref{tab:stress}). The strongest gain perturbation lowered accuracy by about $45\pp$ across the three models. These exploratory contrasts give no evidence that GRN preserved accuracy under the tested perturbations. The reported $p$-values are unadjusted.

\subsection{Masked Reconstruction Pilot on IV-2a}

Motivated by the introduction of GRN in masked image pretraining \cite{convnextv2} and the use of masked reconstruction in EEG pretraining \cite{lcm}, we tested whether a short masked EEG reconstruction warm-up changed the accuracy difference between GRN and the parameter-matched affine bridge. Our $2\times2$ pilot crossed GRN versus affine with masked reconstruction versus supervised warm-up (nine participants, two seeds, 72 runs). Warm-up used 30 epochs on the strict fitting subset. The masked arm zeroed six of fifteen 64-sample blocks and used a shared linear decoder with squared error on masked waveforms. The supervised arm used cross-entropy on unmasked trials. Both arms reset the classifier and optimizer before 60 strict fine-tuning epochs and selected weights on T validation.

Both bridges achieved 67.36\% after supervised warm-up. After masked warm-up, affine and GRN achieved 67.01\% and 67.03\%, respectively. The recipe-by-bridge interaction, defined as GRN minus affine under masking minus the same contrast under supervised warm-up, was $+0.019\pp$ (95\% CI $[-0.23,+0.33]\pp$, raw $p=1.000$). The pilot concerns this short reconstruction and fine-tuning schedule.

\section{Discussion and Conclusion}

The strict comparisons found no detectable accuracy gain from GRN at the tested bridge location. CTNet scores varied across protocols, motivating matched fitting, validation and checkpoint rules. With trained weights fixed, the tested response replacements rarely changed top-1 decisions. The energy-weighted replay illustrates why tests of feature error should also measure decision changes.

Each benchmark has nine participants and evaluates within-participant transfer. The strict IV-2a CI includes a $+0.5\pp$ gain, and the IV-2b upper endpoint is $+0.15\pp$. The sensor tests model selected perturbations, and the masked pilot uses a short schedule. Calibration-specific errors, training dynamics and other backbones remain open. The archive contains code, splits and offline checks for Tables~\ref{tab:main} and~\ref{tab:margin} \cite{materials}.

\clearpage
\nobalance
\section*{Acknowledgment}
OpenAI Codex (GPT-6) assisted in drafting and structuring the abstract and Sections I--V and in producing the vector figures and analysis code. Anthropic Claude Code (Claude Opus 5.5) assisted with scripting and analysis of the benchmark, refit and fixed-response replays in Sections II--IV and checked reported values against the run artifacts. The figures were rendered from the model specification and archived measurements.


\begin{thebibliography}{99}

\bibitem{cross_session}
J. Ma, B. Yang, W. Qiu, Y. Li, S. Gao, and X. Xia, ``A large EEG dataset for studying cross-session variability in motor imagery brain-computer interface,'' \emph{Sci. Data}, vol. 9, Art. no. 531, 2022, doi: \href{https://doi.org/10.1038/s41597-022-01647-1}{10.1038/s41597-022-01647-1}.

\bibitem{eegnet}
V. J. Lawhern, A. J. Solon, N. R. Waytowich, S. M. Gordon, C. P. Hung, and B. J. Lance, ``EEGNet: A compact convolutional neural network for EEG-based brain--computer interfaces,'' \emph{J. Neural Eng.}, vol. 15, no. 5, Art. no. 056013, 2018, doi: \href{https://doi.org/10.1088/1741-2552/aace8c}{10.1088/1741-2552/aace8c}.

\bibitem{qeegnet}
C.-S. Chen, S. Y.-C. Chen, and H.-H. Tseng, ``Exploring the potential of QEEGNet for cross-task and cross-dataset electroencephalography encoding with quantum machine learning,'' \emph{J. Signal Process. Syst.}, vol. 98, Art. no. 5, 2026, doi: \href{https://doi.org/10.1007/s11265-025-01979-2}{10.1007/s11265-025-01979-2}.

\bibitem{conformer}
Y. Song, Q. Zheng, B. Liu, and X. Gao, ``EEG Conformer: Convolutional Transformer for EEG decoding and visualization,'' \emph{IEEE Trans. Neural Syst. Rehabil. Eng.}, vol. 31, pp. 710--719, 2023, doi: \href{https://doi.org/10.1109/TNSRE.2022.3230250}{10.1109/TNSRE.2022.3230250}.

\bibitem{ctnet}
W. Zhao, X. Jiang, B. Zhang, S. Xiao, and S. Weng, ``CTNet: A convolutional transformer network for EEG-based motor imagery classification,'' \emph{Sci. Rep.}, vol. 14, Art. no. 20237, 2024, doi: \href{https://doi.org/10.1038/s41598-024-71118-7}{10.1038/s41598-024-71118-7}.

\bibitem{spd}
C.-S. Chen, E.-J. Kuo, G.-Y. Chen, X. Zhang, and F. Zhang, ``A unified SPD token Transformer framework for EEG classification: Systematic comparison of geometric embeddings,'' \href{https://arxiv.org/abs/2601.21521}{arXiv:2601.21521}, 2026.

\bibitem{convnextv2}
S. Woo, S. Debnath, R. Hu, X. Chen, Z. Liu, I. S. Kweon, and S. Xie, ``ConvNeXt V2: Co-designing and scaling ConvNets with masked autoencoders,'' in \emph{Proc. IEEE/CVF Conf. Comput. Vis. Pattern Recognit.}, 2023, pp. 16133--16142.

\bibitem{maeconformer}
S. Yu, W. P. Marnane, G. B. Boylan, and G. Lightbody, ``Neonatal hypoxic-ischaemic encephalopathy classification from the EEG and HRV signals using a Conformer based masked autoencoder,'' \href{https://arxiv.org/abs/2607.23554}{arXiv:2607.23554}, 2026.

\bibitem{bcic2a}
C. Brunner, R. Leeb, G. R. M\"uller-Putz, A. Schl\"ogl, and G. Pfurtscheller, ``BCI Competition 2008---Graz data set A,'' technical description, 2008. [Online]. Available: \url{https://www.bbci.de/competition/iv/desc_2a.pdf}.

\bibitem{bcic2b}
R. Leeb, C. Brunner, G. R. M\"uller-Putz, A. Schl\"ogl, and G. Pfurtscheller, ``BCI Competition 2008---Graz data set B,'' technical description, 2008. [Online]. Available: \url{https://www.bbci.de/competition/iv/desc_2b.pdf}.

\bibitem{holm}
S. Holm, ``A simple sequentially rejective multiple test procedure,'' \emph{Scand. J. Stat.}, vol. 6, no. 2, pp. 65--70, 1979.

\bibitem{scoring}
T. Gneiting and A. E. Raftery, ``Strictly proper scoring rules, prediction, and estimation,'' \emph{J. Am. Stat. Assoc.}, vol. 102, no. 477, pp. 359--378, 2007, doi: \href{https://doi.org/10.1198/016214506000001437}{10.1198/016214506000001437}.

\bibitem{materials}
CTNet EEG project, ``Paired CTNet and GRN evaluation,'' reproducibility archive, 2026. [Online]. Available at \href{https://github.com/ChiShengChen/ctnet-grn-embc-evidence/tree/15a334e25c2359fbe8ff1c00d324f897a28eb213}{GitHub commit \texttt{15a334e25c23}}. SHA-256 prefix \texttt{a1b49b4db5da}.

\bibitem{lcm}
C.-S. Chen, Y.-J. Chen, and A. H.-W. Tsai, ``Large cognition model: Towards pretrained EEG foundation model,'' \href{https://arxiv.org/abs/2502.17464}{arXiv:2502.17464}, 2025.

\end{thebibliography}
\end{document}